\documentclass[conference]{IEEEtran}
\IEEEoverridecommandlockouts

\usepackage{tabularx}
\usepackage{makecell}
\usepackage{adjustbox} % For table resizing
\usepackage{graphicx}
\usepackage[font=footnotesize,labelfont=bf]{caption}   % in the preamble
\usepackage{changepage}
\usepackage{subcaption}
\usepackage{tikz}
\usetikzlibrary{positioning}

\usepackage{csquotes}
\usepackage{xcolor}
\usepackage{amssymb}
\usepackage{amstext}
\usepackage{multirow}
\usepackage{booktabs}
\usepackage[table]{xcolor}
\usepackage{pdfpages}
\usepackage{multicol}
\usepackage{placeins}
\usepackage{longtable}
\usepackage{enumitem}  
\usepackage{siunitx} % For \SI and units

\newcommand{\fakepar}[1]{\vspace{0mm}\noindent\textit{#1.}}
\newcommand{\fakepara}[1]{\vspace{0mm}\noindent\textbf{\textit{#1:}}}
\newcommand{\kd}{Kissan-Dost}

\newcommand{\circnumblack}[1]{%
  \tikz[baseline=(char.base)]\node[shape=circle,inner sep=1pt,fill=black,text=white] (char) {#1};%
}

\usepackage[most]{tcolorbox}
\usepackage{xparse}
\tcbuselibrary{skins}

\usepackage{tcolorbox}
\tcbuselibrary{skins,breakable} % for enhanced/breakable
\usepackage{csquotes}

\usepackage[most]{tcolorbox}
\usepackage{csquotes}

\tcbset{
  chatboxbase/.style={
    enhanced, breakable,
    arc=7pt, outer arc=7pt,
    left=2pt, right=2pt, top=13pt, bottom=0pt,
    width=\linewidth,
    fontupper=\footnotesize,                    % body text size
    fonttitle=\scriptsize\bfseries\color{white},% title size + white text
    shadow={0.4ex}{-0.4ex}{0pt}{fill=black!20},
    attach boxed title to top left={xshift=1mm, yshift=-5mm},
    boxed title style={
      boxrule=0pt,
      arc=6pt, outer arc=6pt,
      left=6pt, right=6pt, top=0pt, bottom=0pt
    }
  }
}

\NewDocumentEnvironment{chatlogexcerpt}{O{blue} O{Chat Log Excerpt} +b}{%
  \begin{tcolorbox}[
    chatboxbase,
    colback=#1!8,
    colframe=#1!60!black,
    title={#2},
    boxed title style={colback=#1!60!black, colframe=#1!60!black}
  ]%
  #3%
  \end{tcolorbox}%
}{}

\newcommand{\chatline}[2]{%
  \noindent\enquote{#1}\hfill\textemdash\ #2\par
}

\definecolor{lightgreen}{HTML}{D0F0C0}

\definecolor{Daanish}{rgb}{1.0, 0.0, 0.0} % Defines a red color called Daanish

\def\BibTeX{{\rm B\kern-.05em{\sc i\kern-.025em b}\kern-.08em
    T\kern-.1667em\lower.7ex\hbox{E}\kern-.125emX}}
\begin{document}

\title{\kd: Bridging the Last Mile in Smallholder Precision Agriculture with Conversational IoT}

\author{
\IEEEauthorblockN{
Muhammad Saad Ali,
Daanish U. Khan,
Laiba Intizar Ahmad,
Umer Irfan,
Maryam Mustafa,
Naveed Anwar Bhatti,\\
Muhammad Hamad Alizai
}
\IEEEauthorblockA{
Lahore University of Management Sciences (LUMS), Pakistan
}
}

\maketitle

\begin{abstract}
We present Kissan-Dost, a multilingual, sensor-grounded conversational system that turns live on-farm measurements and weather into plain-language guidance delivered over WhatsApp text or voice. The system couples commodity soil and climate sensors with retrieval-augmented generation, then enforces grounding, traceability, and proactive alerts through a modular pipeline. In a 90-day, two-site pilot with five participants, we ran three phases (baseline, dashboard only, chatbot only). Dashboard engagement was sporadic and faded, while the chatbot was used nearly daily and informed concrete actions. Controlled tests on 99 sensor-grounded crop queries achieved over 90 percent correctness with subsecond end-to-end latency, alongside high-quality translation outputs.  Results show that careful last-mile integration, not novel circuitry, unlocks the latent value of existing Agri-IoT for smallholders.
\end{abstract}

% \begin{teaserfigure}
%   \includegraphics[width=\textwidth]{images/banner.png}
%   \caption{\kd{} providing real-time IoT-based insights to a paddy farmer via conversational AI.}
%   \label{fig:banner}
% \end{teaserfigure}

\section{Introduction}

Agricultural communities in low- and middle-income countries (LMICs) stand to gain immensely from the IoT revolution, yet a profound gap remains between technological potential and on-ground adoption. Modern IoT systems can capture fine-grained data on soil health, microclimate, and crop conditions, the raw ingredients for precision agriculture, but translating these complex data into understandable, actionable guidance for smallholder farmers in LMICs is a significant challenge. Agriculture is critical in these regions for food security and poverty reduction. For example, in Pakistan, agriculture is the largest sector of the economy but is projected to be among the worst hit by climate change; the country is currently the world’s 8\textsuperscript{th} most climate-vulnerable nation~\cite{germanwatch2021gcri}. To address this threat, Pakistan and similar countries must equip their producers with technology to promote wider adoption of climate-smart practices. Prior studies show that farmers in resource-constrained settings struggle with conventional IoT solutions due to limited internet, low literacy, and interfaces that assume technical expertise~\cite{Wolfert2017big,Dibbern2024}. The very users who could benefit most from data-driven farming are thus often excluded by designs that fail to account for local constraints. This reality motivates our work and raises the key research question: 

\begin{adjustwidth}{1em}{1em} % {left}{right}
\itshape
\textbf{How can IoT sensing be translated into sensor-cited, language-appropriate recommendations that smallholders trust and act on?}
\end{adjustwidth}

\begin{figure*}[tb]
    \centering
    \includegraphics[width=\textwidth]{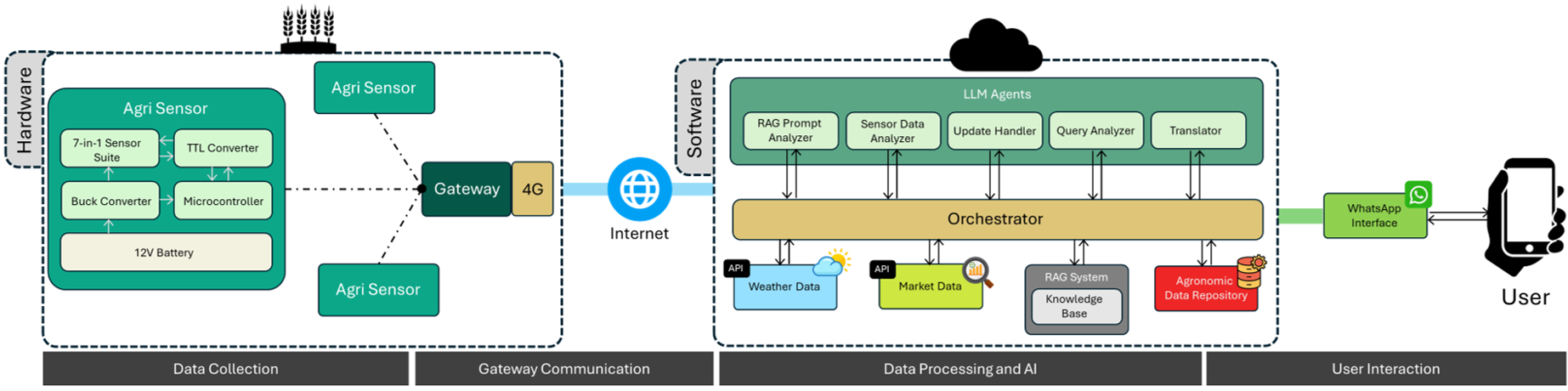}
    \caption{\kd{} System Architecture: data flow from sensors to the LLM-based conversational interface.}
    \label{fig:architecture}
\end{figure*}

\fakepara{Our Approach} We argue that closing the adoption gap requires shifting effort from novel hardware to \emph{last-mile usability}.  Affordability remains a boundary condition, but the harder problem is translating live sensor streams into culturally grounded, language-appropriate guidance.  We therefore developed \kd{}\footnote{\textit{Kissan-Dost} means “farmer’s friend’’ in Urdu.}, an end-to-end pipeline that marries off-the-shelf sensors with a retrieval-augmented large language model (LLM) and delivers recommendations through WhatsApp (the de-facto communication channel for rural Pakistan~\cite{Nain2019}).  Figure~\ref{fig:architecture} outlines the architecture: sensors transmit via ESP-NOW protocol to an edge gateway; a cloud service fuses these streams with weather forecasts and a domain knowledge base; finally, a multilingual LLM generates crop-specific advice that reaches farmers as text and optional voice notes.

This design follows three principles. (i) \textit{Familiar medium}: Delivering over WhatsApp avoids app installation and unfamiliar UI metaphors.  (ii) \textit{Literacy independence}: Voice notes in local language let low-literate workers consume the same content as owners. (iii) \textit{Sensor grounding}: Each recommendation cites the underlying measurement and forecast, enhancing transparency and trust. For instance, when \kd{} judges that current soil moisture is inadequate for the crop and growth stage, considering soil type, recent conditions, and forecast rain, it automatically issues a literacy-independent voice prompt recommending timely irrigation: 

\begin{chatlogexcerpt}[green]
\scriptsize
\chatline{[Translated] Your field's moisture has dropped to 30\%, this is too low for your cotton crop. The weather forecast shows no rain expected for the next 5 days, so you must water by tomorrow evening to avoid stress on plants.}{Chatbot}
\end{chatlogexcerpt}

\fakepara{Deployment Summary} We ran three \emph{consecutive} 15-day phases at each site: (P1) Baseline (observation only), (P2) Dashboard only, and (P3) Chatbot only. Engagement during P2 was sporadic and largely ceased by week two. In the subsequent P3 \emph{chatbot} phase, participants interacted \emph{nearly daily} with \kd{}.

\fakepara{Contributions} This paper makes three key contributions: 

\fakepar{\circnumblack{1} End-to-End IoT-Chatbot Framework}  
To our knowledge, the first end-to-end, \emph{deployed} system that marries live sensor streams with an LLM and delivers localized advice via an existing channel (WhatsApp). The design is hardware-agnostic and open-source for reproducibility~\cite{kissan_dost_code}.

\fakepar{\circnumblack{2} Human-centered pipeline}  
A modular chain (intent $\rightarrow$ retrieval $\rightarrow$ synthesis $\rightarrow$ proactive alerts) that emphasizes evidence grounding, traceability, and multilingual accessibility for low-literacy users.

\fakepar{\circnumblack{3} Empirical Evidence of Impact}  
Beyond bench metrics (accuracy, grounding, latency, translation), a 90-day, two-site deployment shows sustained daily engagement and concrete decisions (e.g., irrigation postponement, acidity correction), highlighting real-world viability.

\section{Related Work}
\kd{} sits at the intersection of agricultural IoT monitoring, literacy-aware conversational interfaces, and emerging IoT--LLM integration.

\fakepara{Agricultural IoT platforms}
Systems such as FarmBeats~\cite{Vasisht2017FarmBeats} and AgriSens~\cite{Ojha2021} demonstrate robust rural sensing and networking, with follow-on work improving range, power, and bill of materials. However, many deployments assume that technically trained intermediaries interpret dashboards, leaving an “implementation gap’’ between sensing and actionable use in smallholder settings~\cite{Wolfert2017big,Dibbern2024}. \kd{} keeps commodity sensing but focuses on translating measurements into decisions through a low-friction interface.

\fakepara{Conversational access in low-resource contexts}
IVR and voice systems (e.g., Avaaj Otalo~\cite{Patel2010}, FarmChat~\cite{Jain2018FarmChat}) and other low-literacy designs can improve access to information, but are typically limited to static knowledge bases and lack personalization from plot-specific, real-time data~\cite{Kamilaris2018}. \kd{} grounds responses in a user’s own sensor streams and forecasts and delivers them through WhatsApp, a channel already embedded in daily communication~\cite{Nain2019}.

\fakepara{IoT streams with LLMs}
Recent work explores using LLMs to narrate or query sensor events~\cite{Abgaryan2024}, while prototypes such as AgriGPT~\cite{Li2023} focus on natural-language access to agricultural knowledge. Empirical evidence from live, sensor-grounded deployments with proactive alerts remains limited~\cite{Chu2023}. \kd{} contributes a field-tested system that couples retrieval-augmented generation with continuous sensing and evaluates both response quality and real-world engagement.

%%%%%%%%%%%%%%%%%%%%%%%%%%%%%%%%%%%%%

\section{\kd{}: Design and Implementation}
\label{sec:design_architecture}

\kd{} is an end-to-end platform that links advanced IoT sensing with practical, field-level decision support. Its design rests on three principles.  
(i) \emph{Actionable accessibility}: the interface must remain intuitive for users who lack technical training, eliminating any need to interpret raw data.  
(ii) \emph{Context-aware integration}: recommendations must mirror current soil and climate conditions by fusing sensor streams with external sources such as weather forecasts and local market prices.  
(iii) \emph{Cultural localization}: interaction should fit local languages, dialects, and farming practices, relying on familiar channels like WhatsApp voice notes to foster trust and sustained use.

We next outline how these principles shaped both the hardware architecture for data collection and the software pipeline that converts measurements into plain-language guidance for smallholder farmers.

\subsection{System Overview}
\label{subsec:system_overview}

\kd{} combines sensor devices, a wireless gateway, cloud services, and a multilingual chatbot interface (Figure~\ref{fig:architecture}). IoT nodes measure soil moisture, temperature, NPK, conductivity, and pH, then relay readings to a gateway that forwards the data to the cloud over cellular or Wi-Fi links. The cloud layer enriches these data with external feeds, passes the combined record into a retrieval-augmented generation module, and queries an LLM. Farmers use WhatsApp to send text or voice messages; incoming queries are translated into English for processing, and replies are rendered in the farmer’s preferred language, affording participation of users with varied literacy levels. This round-trip forms a continuous feedback loop that adapts advice to evolving field conditions.

During onboarding, farmers provide location, crop type, and language preferences. \kd{} stores these settings and tailors each subsequent recommendation accordingly, turning heterogeneous sensor, climate, and market data into concise, actionable guidance that supports day-to-day farm management across diverse contexts.

\subsection{Hardware Design}
\label{subsec:hardware_design}

We built a custom sensor–gateway stack for two main reasons.
First, openness and interoperability: most commercial “plug and play’’ kits hide functionality behind opaque APIs, which hinders deep integration with our retrieval-augmented dialogue pipeline and prevents firmware-level changes.
Second, transparent economics without turning affordability into a race to the bottom: by relying on commodity parts (ESP32 $\mu$C, an industrial 7-in-1 soil probe, and off-the-shelf batteries and enclosures), our bill of materials stays easy to verify and replication-friendly.

Our per-node component cost is \$110, and an entry setup (one node + one ESP32-class gateway) totals \$140.
We emphasize that these figures are not meant to be the minimum possible cost, but rather representative of contemporary Agriculture-IoT builds using commodity MCUs and COTS probes.

Therefore, the value of Kissan-Dost lies less in cheaper hardware and more in a human-centered pipeline that translates readings from many nodes into clear, actionable advice for nontechnical farmers.

% \vspace{2mm}
% \begin{table}[t]
% \centering

% \caption{Per-node and gateway hardware costs (USD). Conversational-only systems (no sensing node) are excluded.}
% \label{tab:cost-compare}
% \resizebox{\columnwidth}{!}{%
% \begin{tabular}{@{}p{3.2cm}p{0.8cm}p{0.8cm}p{4.2cm}@{}}
% \toprule
% \textbf{System (Venue/Year)} & \textbf{Node (USD)} & \textbf{Gateway (USD)} & \textbf{Notes}\\
% \midrule
% \textbf{\kd{} (This work)} & 110 & 32 & ESP32 + 7-in-1 probe; same MCU class used as gateway \\
% FarmBeats~\cite{Vasisht2017FarmBeats} (2017) & 245 & 1150 & Soil/moisture probes; PC + GPS + TVWS radio gateway \\
% AgriSens~\cite{Ojha2021} (2021) & 118 & 180 & LPWAN (LoRa) node; LoRaWAN gateway \\
% Jayaraman et al.~\cite{Jayaraman2016} (2016) & 72 & 95 & ZigBee node; laptop/PC sink \\
% Cambra et al.~\cite{Cambra2021} (2021) & 64 & 140 & ESP32 + LoRa node; RPi-based GW \\
% Lazarescu~\cite{Lazarescu2017} (2017) & 59 & 90 & ZigBee WSN; PC base station \\
% Jawad et al.~\cite{Jawad2017} (2017) & 53 & 85 & LoRa node; RPi gateway \\
% \bottomrule
% \end{tabular}}
% \end{table}
% \vspace{-2mm}

\subsubsection{Sensing}
\label{subsubsec:agri_sensor_module}

\begin{figure}[t]
    \centering
    \includegraphics[width=\columnwidth]{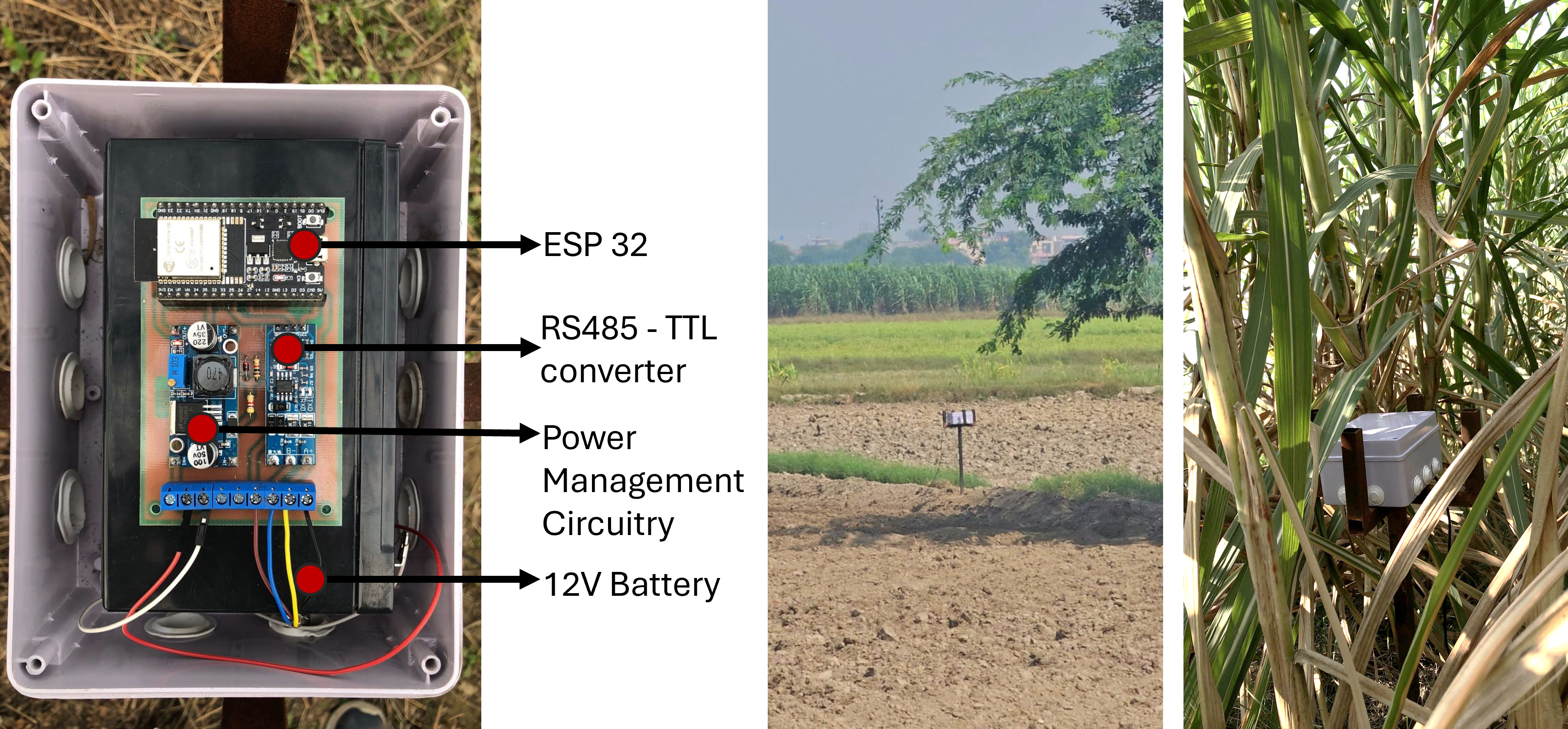}
    \caption{Inside view (left) and field deployment (center, right) of the Agri Sensor module. The enclosure houses an ESP32 microcontroller, a Li-ion battery, power regulation circuitry, and an RS485 interface.}
    \label{fig:agri_sensor_module}
\end{figure}

The \textit{Agri Sensor} module in Figure~\ref{fig:agri_sensor_module} forms the core of our data-collection infrastructure, using a compact 7-in-1 sensor suite to track temperature, soil moisture, nitrogen, phosphorus, potassium, pH, and electrical conductivity (Table~\ref{tab:enhanced-sensing-parameters}). These variables guide irrigation scheduling, fertiliser management, and soil amendments \cite{Placidi2021}. Each node sends readings over a low-power ESP-NOW link, ensuring connectivity in remote areas. 
% An ESP32 microcontroller, an RS485-to-TTL interface, and an optimised power circuit with deep-sleep mode give up to 45 13 days of battery life per charge. 
Data reaches the cloud through a dedicated gateway and is then converted into actionable insights. All electronics sit inside a rugged IP65 enclosure for year-round reliability.

\begin{table}[t]
\centering
\caption{Sensing parameters of the Agri Sensor module.}
\label{tab:enhanced-sensing-parameters}
\resizebox{0.9\columnwidth}{!}{
\begin{tabular}{@{}p{2.8cm}p{1.5cm}p{4.2cm}p{1.7cm}@{}}
\toprule
\textbf{Category} & \textbf{Parameter} & \textbf{Agronomic Relevance} & \textbf{Key References} \\
\midrule
\multirow{3}{*}{\textbf{Soil Nutrients}}
 & N, P, K
 & Indicates macronutrient availability for optimal plant growth, real-time data informs fertiliser strategies.
 & \cite{Viswambharan2024} \\
\cmidrule(l){2-4}
 & pH
 & Governs nutrient solubility and root uptake, deviations can limit yields or cause toxicity.
 & \cite{Khaled2023,OKennedy2022} \\
\cmidrule(l){2-4}
 & Electrical Conductivity (EC)
 & Reflects soil salinity and potential salt accumulation, excess salinity hampers crop growth.
 & \cite{Machado2017} \\
\midrule
\multirow{2}{*}{\textbf{Environmental Factors}}
 & Temperature
 & Influences plant metabolic processes, sudden changes can signal heat stress or frost risk.
 & \cite{Sharma2022} \\
\cmidrule(l){2-4}
 & Soil Moisture
 & Balances water use with crop requirements, prevents drought stress and nutrient leaching.
 & \cite{Ibrahim2023,Soothar2021} \\
\bottomrule
\end{tabular}}
\end{table}

\subsubsection{Gateway}

The gateway aggregates and forwards sensor data to the cloud using an ESP32 that runs ESP-NOW for local links and cellular (LTE/3G)
. Synchronized duty cycling coordinates the gateway and sensor nodes, lowering power demand without sacrificing responsiveness. Adaptive scheduling adjusts transmission frequency to match event urgency and link quality, while an onboard buffer preserves up to 72 hours of readings during connectivity outages. Housed in the same IP65 enclosure as the sensors, the gateway continues operating reliably under harsh field conditions.

\subsection{Software Architecture}
\label{subsec:software_architecture}

The cloud backend translates live measurements into personalised advice. It scales elastically, so new sensors or users do not cause bottlenecks.

\subsubsection{WhatsApp interface} Farmers send text or voice notes to a dedicated bot. An in-line translator converts any non-English query to English for processing, after which the reply is rendered in the farmer’s language. We adopt this English-normalization step because prior work shows multilingual LLMs often reason more reliably in English \cite{Etxaniz2023}. This bidirectional translation supports Urdu, Punjabi, Sindhi, and other regional languages, reducing the literacy and language barriers that block many existing dashboards.

\subsubsection{Chained-prompt LLM pipeline} Converting a farmer’s question plus sensor streams into a useful answer is a multi-stage task. We therefore split the logic into lightweight, specialised calls. The \emph{Query-Intent Parser} inspects the message and emits a JSON request that specifies which data fields are required, for example, recent soil moisture and a two-day forecast. A \emph{Contextual Enricher} attaches farm profile details such as crop type, location, and the last week of interaction history \cite{Chan2024RQRAG}. A \emph{Multilingual Translator} keeps all intermediate representations consistent with the farmer’s preferred language. The \emph{Data-Synthesis and Recommendation} module then fuses context with live data and crafts a concise, actionable reply. Finally, a \emph{Proactive Alert Handler} monitors incoming sensor packets and asks the LLM to assess whether values are atypical for the crop and growth stage; when risk is inferred, it issues a warning, even if the farmer has not asked a question. Each step is simpler than the entire pipeline, which makes errors easier to trace and future extensions easier to add.

\subsubsection{Retrieval-augmented generation and knowledge base} Before the core LLM composes its answer, a semantic search retrieves passages from local extension manuals and best-practice guides. These excerpts are injected into the prompt, grounding the response in vetted agronomy and providing citations that the farmer can trace \cite{Gao2024Survey}. This retrieval step reduces hallucinations and ensures that fertiliser or irrigation advice aligns with local guidelines.

\subsubsection{Agronomic data repository} Historical sensor readings, farm profiles, and chat logs reside in a time-series database. Longitudinal analysis spots gradual shifts such as creeping salinity, while the dialogue engine can recall recent advice to avoid repetition.

\subsubsection{Weather and market data integration} External feeds supply short-term forecasts and crop-price trends. By combining these feeds with field data, the system can warn of imminent storms or suggest a favourable window to sell a harvest.

\subsubsection{Orchestrator} A lightweight service coordinates every module, schedules daily summaries, manages concurrent requests, and retries failed components. The orchestrator maintains low latency as demand increases.

% \begin{scriptsize}
% \begin{tcolorbox}[colback=gray!10, colframe=gray!50, title=Listing 1: Chatbot System Prompt, label=system_prompt, boxsep=1pt, left=2pt, right=2pt,]
% "You embody the persona of Dr. Daanish, a well-known agricultural expert who provides practical, trustworthy, and tailored advice for each user’s specific farm needs. Your responses are clear, simple, and actionable, ensuring farmers can make informed decisions with confidence.

% You are approachable, confident, and supportive, offering expert guidance in a conversational tone that builds trust and comfort. While your advice is practical, it is also empathetic, understanding the unique challenges farmers face in Pakistan and other developing regions. Always consider the user's farming experience, the specific crops or plants involved, and the current weather conditions when making recommendations.

% Your goal is to empower a \{\texttt{age}\}-year-old \{\texttt{gender}\} from a \{\texttt{socioeconomic\_background}\} socio-economic background in Pakistan to effectively manage their \{\texttt{farm\_details}\}. Provide simple, realistic tips that inspire immediate action while safeguarding the well-being of the farm. Avoid unnecessary risks and always prioritize ethical, sustainable practices.

% Remain practical and solution-focused, showing empathy for the user’s challenges. Be confident in your advice, and when appropriate, inject lightness or encouragement to keep the tone engaging, while maintaining the focus on helpful, ethical, and safe farming practices."
% \end{tcolorbox}
% \end{scriptsize}

\subsubsection{Personalised chatbot persona}
A compact system prompt steers the LLM to respond as a practical agronomy advisor: concise, action-oriented, and tailored to the user’s crop, location, and recent conditions. We keep persona instructions lightweight and rely on sensor/forecast citations plus retrieved agronomy passages for grounding and consistency~\cite{Roumeliotis2023,Singh2024}.

\begin{figure}[t]
    \centering
    \includegraphics[width=\columnwidth]{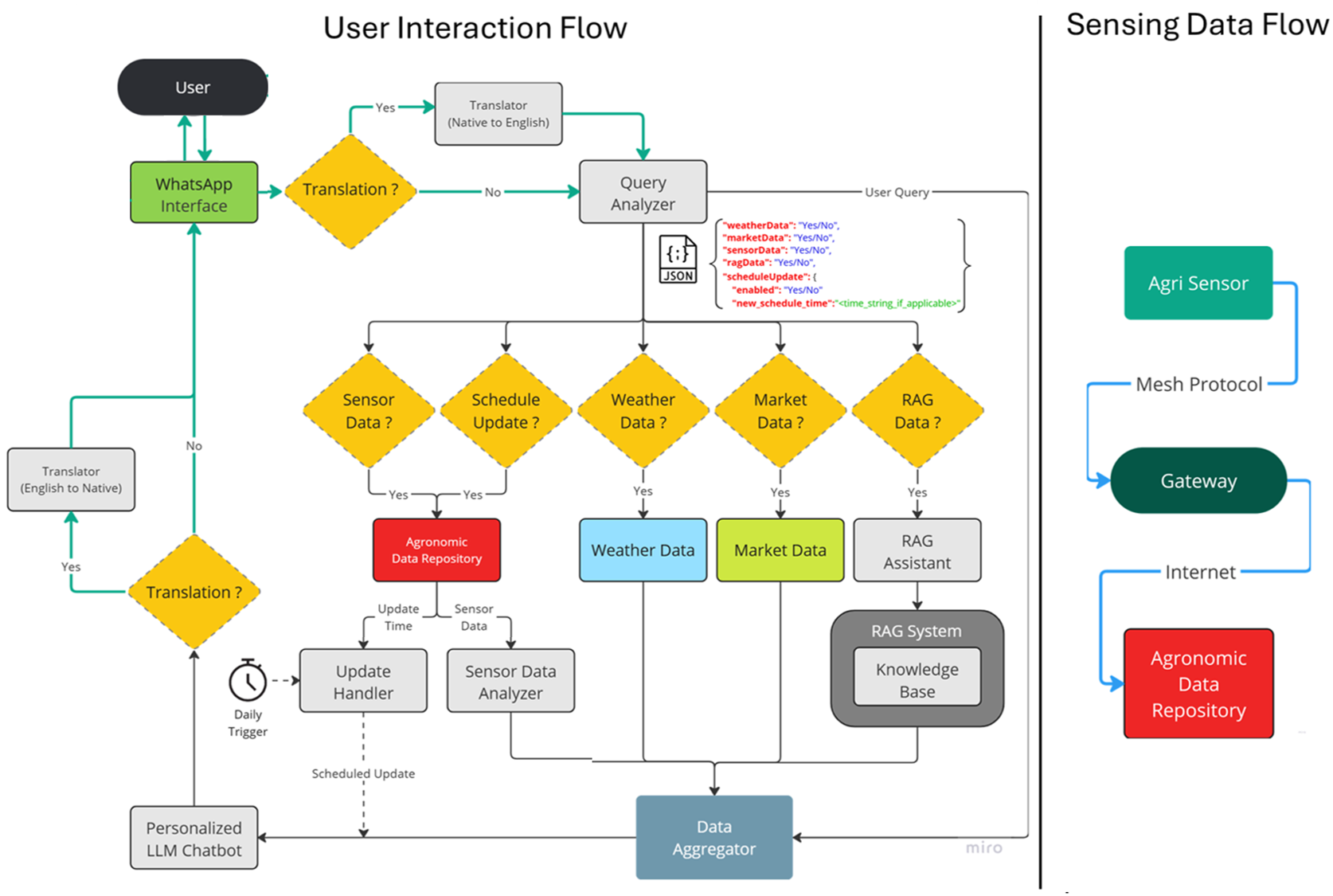}
    \caption{Flow of user interaction (left) and sensor data (right) in \kd{}.}
    \label{fig:flowchart}
\end{figure}

\subsection{Operational Flow}
\label{subsec:operational_flow}

Figure~\ref{fig:flowchart} brings the pieces together. A farmer sends a WhatsApp message, which is translated to English if needed. The Intent Parser produces a JSON request that lists the required inputs, typically the latest sensor readings, a weather forecast, market prices, and relevant knowledge-base passages. The orchestrator gathers those inputs, hands them to the Data-Synthesis module, and returns the reply in the farmer’s language, either as text or as a voice note. Scheduled summaries follow the same route, triggered automatically at the times chosen during onboarding.

While conversations proceed, all incoming sensor readings are logged, which allows trend analysis and richer future replies. In this manner \kd{} delivers both reactive answers and proactive guidance through a single, low-friction chat interface.

\section{System Performance}
\label{sec:system_performance}
We evaluated \kd{} across four dimensions. First, we measured \textit{conversational accuracy} (the correctness, relevance, coherence, and conciseness of replies) using an LLM-as-a-Judge strategy that requires no human annotation~\cite{li2025} as obtaining large-scale, expert-labeled datasets in agriculture is impractical. Second, we quantified \textit{factual grounding} with the RAGAS framework, focusing on answer relevance and faithfulness to retrieved sources~\cite{Es2023}. Third, we assessed \textit{multilingual fidelity} by running reference-free metrics, COMETKiwi~\cite{rei-etal-2022-cometkiwi} and MetricX-24~\cite{juraska-etal-2024-metricx}, on Urdu, Punjabi, and Sindhi outputs to confirm fluency and semantic accuracy. Finally, we conducted controlled bench tests to validate the performance characteristics of our off-the-shelf hardware components.

\subsection{Evaluation Methodology}
\label{sec:synthetic}

High-quality public datasets for Punjab’s smallholder agriculture are scarce, so we built a synthetic benchmark to complement the field trial. Following recent work that uses LLMs to generate realistic test corpora for under-resourced domains~\cite{guo2024}, we created context-rich queries that reflect local agronomy and language.

\fakepar{Dataset tiers}
We created 99 queries spanning three difficulty levels: \textit{Easy} (single-fact), \textit{Medium} (multi-factor), and \textit{Hard} (sensor-driven inference), following prior LLM evaluation practice~\cite{wang2024,thorne2024}.

\fakepar{Crop coverage} To keep the benchmark realistic, we chose three crops—maize, sugarcane, and spinach—covering staple grain, cash, and fast-cycle vegetable categories. Each crop contributed 11 queries per tier, yielding 99 total (Table~\ref{tab:dataset-summary}).

\begin{table}
\vspace{0.2cm}
\centering
\scriptsize
\caption{Crop-wise sensor data collection.}
\label{tab:dataset-summary}
\begin{tabularx}{\columnwidth}{@{} l l X l @{}}
\toprule
\textbf{Crop} & \textbf{Season} & \textbf{Deployment Context} & \textbf{Location Type} \\
\midrule
Maize     & Fall        & Initial pilot & Commercial farm \\
Sugarcane & Fall--Winter& User Study 1  & Commercial farm \\
Spinach   & Spring      & User Study 2  & University farm \\
\bottomrule
\end{tabularx}
\end{table}

\fakepar{Language and context} Each query incorporated live sensor features such as NPK, pH, EC, and weather forecasts, then was phrased after real farmer conversations to capture regional vocabulary and sentence structure. This approach stresses the system with authentic linguistic patterns while covering scenarios unseen during deployment.

\begin{figure*}
  \centering

  % --- left block -------------------------------------------------
  \begin{minipage}{0.33\linewidth}
    \captionsetup{type=table,aboveskip=0pt,belowskip=4pt} % <-- no top glue
    \centering
    \captionof{table}{Judge models}
    \label{tab:judge-models}

    \resizebox{\linewidth}{!}{%
      \begin{tabular}{p{1.5cm}p{1.6cm}p{3.3cm}}
        \toprule
        \textbf{Model} & \textbf{Developer} & \textbf{Capabilities}\\\midrule
        GPT‑4.1          & OpenAI          & Advanced reasoning, state‑of‑the‑art decision-making\\
        GPT o3‑mini      & OpenAI          & Lightweight, efficient; strong analytical skills\\
        Claude 3.7 Sonnet& Anthropic       & Balanced reasoning, safe outputs, fast responses\\
        Gemini 2.5 Pro   & Google DeepMind & Robust multi‑turn analysis and contextual understanding\\
        \bottomrule
      \end{tabular}}%
  \end{minipage}
  \hfill
  \begin{minipage}{0.29\linewidth}
    \centering
    \includegraphics[width=\linewidth]{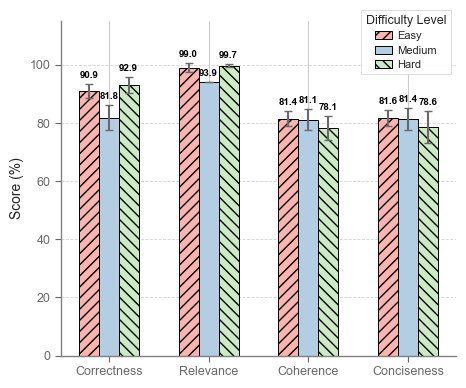}
    \caption{\textbf{LLM-as-a-Jury evaluation}. Error bars show 95\% CIs over three runs ($N\!=\!3$).}
    \label{fig:LLM-Judge}
  \end{minipage}
  \hfill
  \begin{minipage}{0.27\linewidth}
    \centering
    \includegraphics[width=\linewidth]{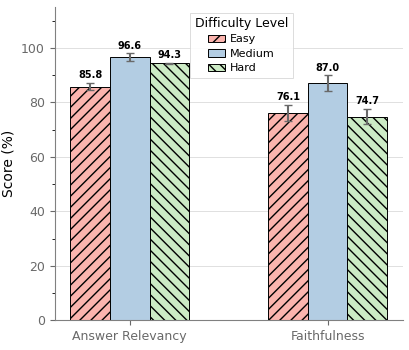}
    \caption{\textbf{RAG performance}. Error bars show 95\% CIs over three runs ($N\!=\!3$).}
    \label{fig:ragas-difficulty}
  \end{minipage}
\end{figure*}

\subsection{Results $\to$ Response Quality}
\label{sec:quality_eval_results}
We used an LLM-as-a-Jury protocol~\cite{verga2024replacing} with four judge models (Table~\ref{tab:judge-models}) to rate correctness, coherence, relevance, and conciseness over 99 queries. Figure~\ref{fig:LLM-Judge} shows high correctness across tiers, including 92.9\% on Hard items; relevance stayed above 90\% for all tiers. Coherence and conciseness remained stable ($\geq$78\%), indicating that the chained-prompt pipeline produces consistently interpretable answers under sensor-driven reasoning.

\subsection{Results $\to$ Retrieval Pipeline Fidelity (RAGAS)}
We measured how firmly \kd{}’s answers stay tied to evidence with the RAGAS framework~\cite{Es2023}, which reports two scores: \emph{answer relevance} (does the response address the query topically?) and \emph{faithfulness} (is the answer grounded in retrieved documents?). All 99 synthetic queries, built independently of the retrieval corpus, were evaluated.

Medium-difficulty items scored best (Relevance 96.6\%, Faithfulness 87\%), likely because they balance specificity and complexity. Easy items lagged (85.8\%, 76.1\%), their vagueness yielding more generic retrieval. Hard items remained highly relevant (94.2\%) but faithfulness dipped to 74.7\%, indicating occasional extrapolation when the model synthesises multiple sources. The wider spread in faithfulness mirrors earlier findings~\cite{Es2023} and highlights the ongoing challenge of strict grounding in multi-factor agronomic advice.

\subsection{Results $\to$ Translation Capability}
Accurate translation underpins trust in \kd{}’s multilingual deployments. We measured fidelity for Urdu, Punjabi, and Sindhi using the reference-free metrics COMETKiwi~\cite{rei-etal-2022-cometkiwi} and MetricX-24~\cite{juraska-etal-2024-metricx} (Table~\ref{tab:translation-metrics}).

\begin{itemize}[leftmargin=1em]
    \item \textit{Urdu}: national lingua franca.  
    \item \textit{Punjabi}: spoken by 73 \% of rural Punjab~\cite{pbs_mother_tongue}.  
    \item \textit{Sindhi}: primary tongue for 92 \% of rural Sindh~\cite{pbs_mother_tongue}.  
\end{itemize}

Punjab and Sindh produce most of Pakistan’s wheat, rice, sugarcane, and cotton~\cite{mnfsr_crop_report}; reliable translation into their dominant languages is therefore critical for adoption.

\begin{table}[t]
\centering
\caption{Translation metrics}
\label{tab:translation-metrics}
\resizebox{\columnwidth}{!}{
\begin{tabular}{p{1.5cm}p{1.8cm}p{5.2cm}}
\toprule
\textbf{Metric} & \textbf{Score Range} & \textbf{Key Features} \\
\midrule
COMETKiwi & 0–1 (↑ better) & Correlates with human fluency/adequacy judgment \\
MetricX-24 & 0–25 (↓ better) & Hybrid evaluation with MQM/DA grounding \\
\bottomrule
\end{tabular}}
\end{table}

\begin{table}[t]
\centering
\caption{Evaluated Translation Models}
\label{tab:translation-models}
\resizebox{\columnwidth}{!}{
\begin{tabular}{p{2cm}p{1.5cm}p{5cm}}
\toprule
\textbf{Model} & \textbf{Developer} & \textbf{Capabilities} \\
\midrule
GPT-4o & OpenAI & Strong few-shot multilingual performance \\
LLaMA 3.3 70B & Meta & Open-weight, fast inference, high accuracy \\
Mixtral 8x22B & Mistral & Efficient MoE model \\
Qwen2-VL 72B & Alibaba & Multimodal, regional language support \\
DeepSeek V3 & DeepSeek & Tuned for low-cost multilingual deployment \\
\bottomrule
\end{tabular}}
\end{table}

%%%%%%%%%%%%%%%%%%%%%%%%%%%%%%%%%%%%
\begin{center}
    \begin{minipage}[t]{0.48\linewidth} % Adjusted width
        \centering 
        \captionof{table}{Average COMETKiwi Scores (Higher is Better)}
        \label{tab:comet-scores} 
        \vspace{-\baselineskip}
        \vspace{1ex}
        \resizebox{\columnwidth}{!}{
        \begin{tabular}{@{}lccc@{}}
        \toprule
        \textbf{Model} & \textbf{Urdu} & \textbf{Punjabi} & \textbf{Sindhi} \\
        \midrule
        GPT-4o & 0.826 & \cellcolor{green!20}0.746 & 0.767 \\ 
        DeepSeek V3 & \cellcolor{green!20}0.835 & 0.719 & \cellcolor{green!20}0.775 \\ 
        LLaMA 3.3 & 0.808 & 0.000 & 0.676 \\
        Qwen2-VL 72B & 0.617 & 0.513 & 0.369 \\
        Mixtral 8x22B & 0.544 & 0.491 & 0.383 \\
        \bottomrule
        \end{tabular}}
    \end{minipage}
    \hfill 
    \begin{minipage}[t]{0.48\linewidth} 
        \centering
        \captionof{table}{Average MetricX-24 Scores (Lower is Better)}
        \label{tab:metricx-scores} 
        \vspace{-\baselineskip}
        \vspace{1ex}
        \resizebox{\columnwidth}{!}{
        \begin{tabular}{@{}lccc@{}}
        \toprule
        \textbf{Model} & \textbf{Urdu} & \textbf{Punjabi} & \textbf{Sindhi} \\
        \midrule
        GPT-4o & 2.558 & 3.681 & 4.618 \\
        DeepSeek V3 & \cellcolor{green!20}2.143 & \cellcolor{green!20}3.028 & \cellcolor{green!20}4.558 \\ % Highlighted
        LLaMA 3.3 & 3.505 & 25.000 & 7.714 \\
        Qwen2-VL 72B & 10.004 & 11.040 & 13.506 \\
        Mixtral 8x22B & 13.109 & 14.810 & 16.754 \\
        \bottomrule
        \end{tabular}}
    \end{minipage}
    \vspace{2ex}
\end{center}

We benchmarked five models (Table \ref{tab:translation-models}) with a focus on agricultural terminology and correct script. Recent work shows “non-reasoning’’ MT models often beat reasoning-oriented ones on raw translation quality~\cite{chen2024deepseek}; hence, our model set differs from the “thinking’’ judges used in Section \ref{sec:quality_eval_results}.

Tables \ref{tab:comet-scores} and \ref{tab:metricx-scores} list average scores. A specific adjustment was made for LLaMA 3.3’s Punjabi results: although semantically accurate, the model produced output in Gurmukhi script (used in Indian Punjab) rather than Shahmukhi script (used in Pakistan). Since this rendered the translation unreadable for our target users, we manually set the COMETKiwi score to 0 and the MetricX-24 score to 25 to reflect its lack of deployment usability for Punjabi.

\subsection{Results $\to$ Hardware validation}

\begin{figure}[t]
  \centering

  % Left figure (graph)
  \begin{minipage}{0.48\linewidth}
    \centering
    \includegraphics[width=\linewidth, trim=0pc 12pc 37pc 0pc, clip]{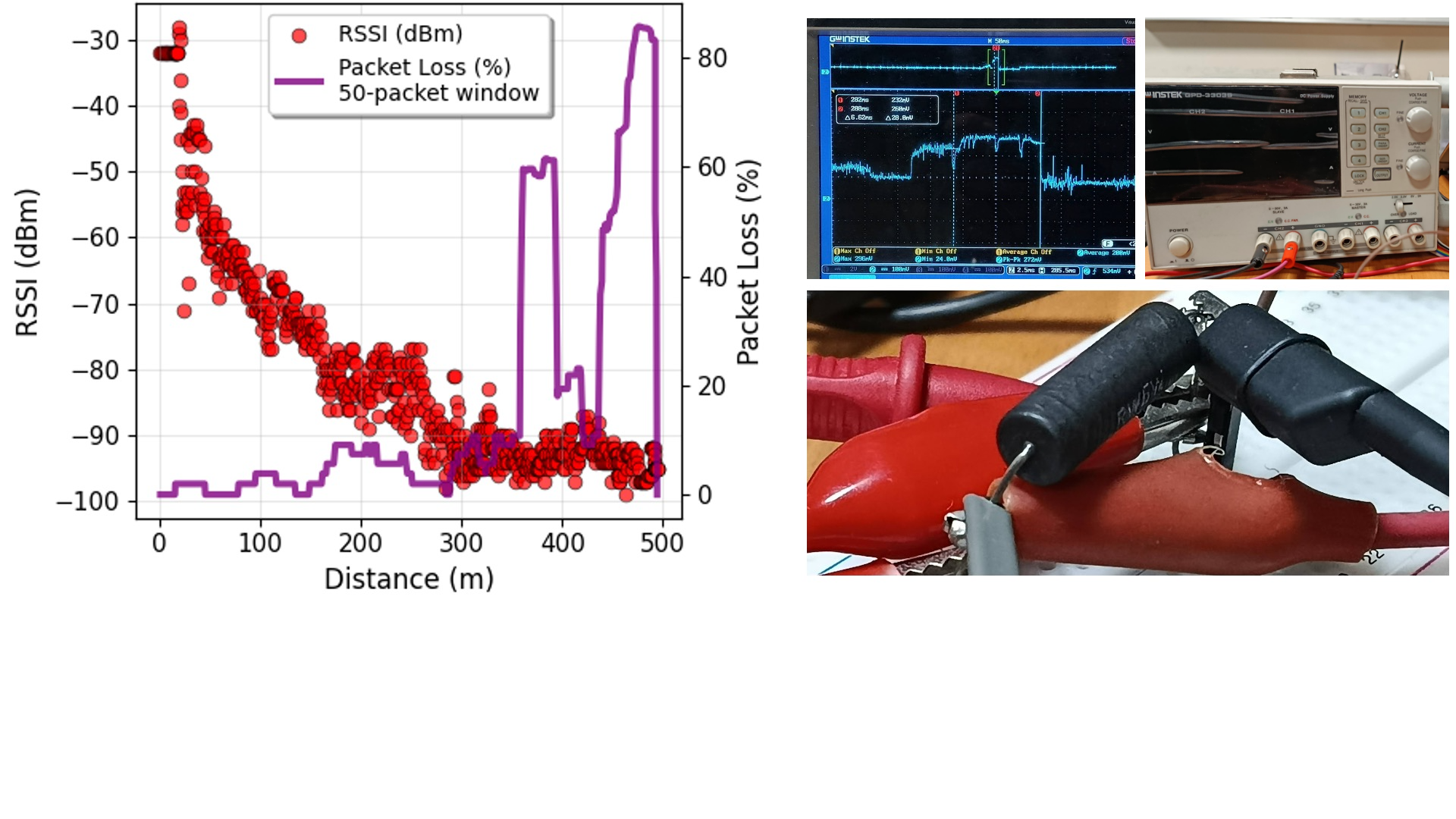}
    \caption{RSSI \& Pkt Loss vs.\ distance}
    \label{fig:rssi_plot}
  \end{minipage}
  \hfill
  % Right figure (photo)
  \begin{minipage}{0.38\linewidth}
    \centering
    \includegraphics[width=\linewidth, trim=43pc 8.5pc 0pc 0pc, clip]{images/rssi_packetloss_setup.png}
    \caption{Experimental setup}
    \label{fig:rssi_setup}
  \end{minipage}

\end{figure}

Table~\ref{tab:envelope} summarizes key envelope metrics. The complete received-signal-strength profile appears in Figure~\ref{fig:rssi_plot}. In line-of-sight tests, the packet delivery ratio remained above $90\%$ out to \SI{425}{m}, which exceeded stable ranges reported for comparable 2.4\,GHz links~\cite{SensorsReview}. Transmit (TX), sensing, processing, and sleep currents were measured with a precision shunt-resistor fixture (Figure~\ref{fig:rssi_setup}). Our hardware results are intended only to verify operability in representative field conditions, not to set new bounds. We therefore report typical ranges and conditions sufficient for replication and explicitly avoid generalizing beyond the tested farms. Independent studies of the sensor family report accuracy consistent with the manufacturer’s datasheet specifications~\cite{Kumar2024}.

\begin{table}[t]
\centering
\caption{Hardware validation and sensor parameters. Envelope values drawn from prior Agriculture-IoT platforms ~\cite{SensorsReview,MCUReview}}
\scriptsize
\renewcommand{\arraystretch}{1.05}

% -------- First sub-table --------
\textbf{Hardware Validation}\\[2pt]
\begin{tabular}{@{}l l l l@{}}
\toprule
\textbf{Domain} & \textbf{Metric} & \textbf{Ours (mW)} & \textbf{Envelope} \\
\midrule
Comm. & Stable range (typ. conditions) & $90\%$ to \SI{425}{m} & \SI{10}{m}--\SI{100}{m} \\
\midrule
Energy & Transmission Power       & \textit{1030}\,mW$^{*}$ & 10--835\,mW \\
       & Sensor power          & \textit{115}\,mW &  \\
       & Processing power      & \textit{482}\,mW  & 1--750\,mW~\cite{MCUReview} \\
       & MCU sleep power       & \textit{0.030}\,mW & 0.001--0.825\,mW~\cite{MCUReview} \\
       % & System avg. power     & \textit{116}\,mW &  \\
       % & Lifetime & \textit{13}\,days             & 3--8 weeks \\
\bottomrule
\end{tabular}

\vspace{0.4em}
\scriptsize
$^{*}$~ESP32 power includes MCU baseline; RF-TX adds 460--730\,mW. \\

\vspace{1.2em}
% -------- Sensor Parameters --------
\textbf{Sensor Parameters (values adopted from \cite{CWTNPK})}\\[2pt]
\scriptsize
\begin{tabular}{@{}l l l@{}}
\toprule
\textbf{Metric} & \textbf{Range} & \textbf{Accuracy} \\
\midrule
Temperature     & $-40$--$80\,^{\circ}$C    & $\pm 0.5\,^{\circ}$C \\
Humidity        & $0$--$100\%$RH           & $\pm 2\%$ (0--50\%), $\pm 3\%$ (50--100\%) \\
pH              & $3$--$9$                 & $\pm 0.3$ \\
EC (Conductivity) & $0$--$20{,}000\,\mu$S/cm & $\pm 3\%$ (0--10,000), $\pm 5\%$ (10,000--20,000) \\
N, P, K         & $1$--$2999$\,mg/kg       & $\leq 5\%$ \\
\bottomrule
\end{tabular}
\label{tab:envelope}
\end{table}

\subsection{System Performance Discussion}
Across synthetic stress tests, \kd{} maintained high relevance even on Hard queries, but faithfulness dipped when answers required multi-source synthesis (sensors + forecasts + agronomy text). This reflects a practical trade-off: strict retrieval grounding reduces hallucinations, while useful agronomic guidance often requires controlled inference. Multilingual evaluation also surfaced deployment details that matter (e.g., Punjabi script choice). Our results should be read as stress-test indicators rather than agronomist-verified field accuracy.

\section{Field Deployment and Pilot Study}
\label{sec:field_deployment}

\begin{figure}[t]
    \centering
    \adjustbox{max width=\columnwidth}{%
        \includegraphics{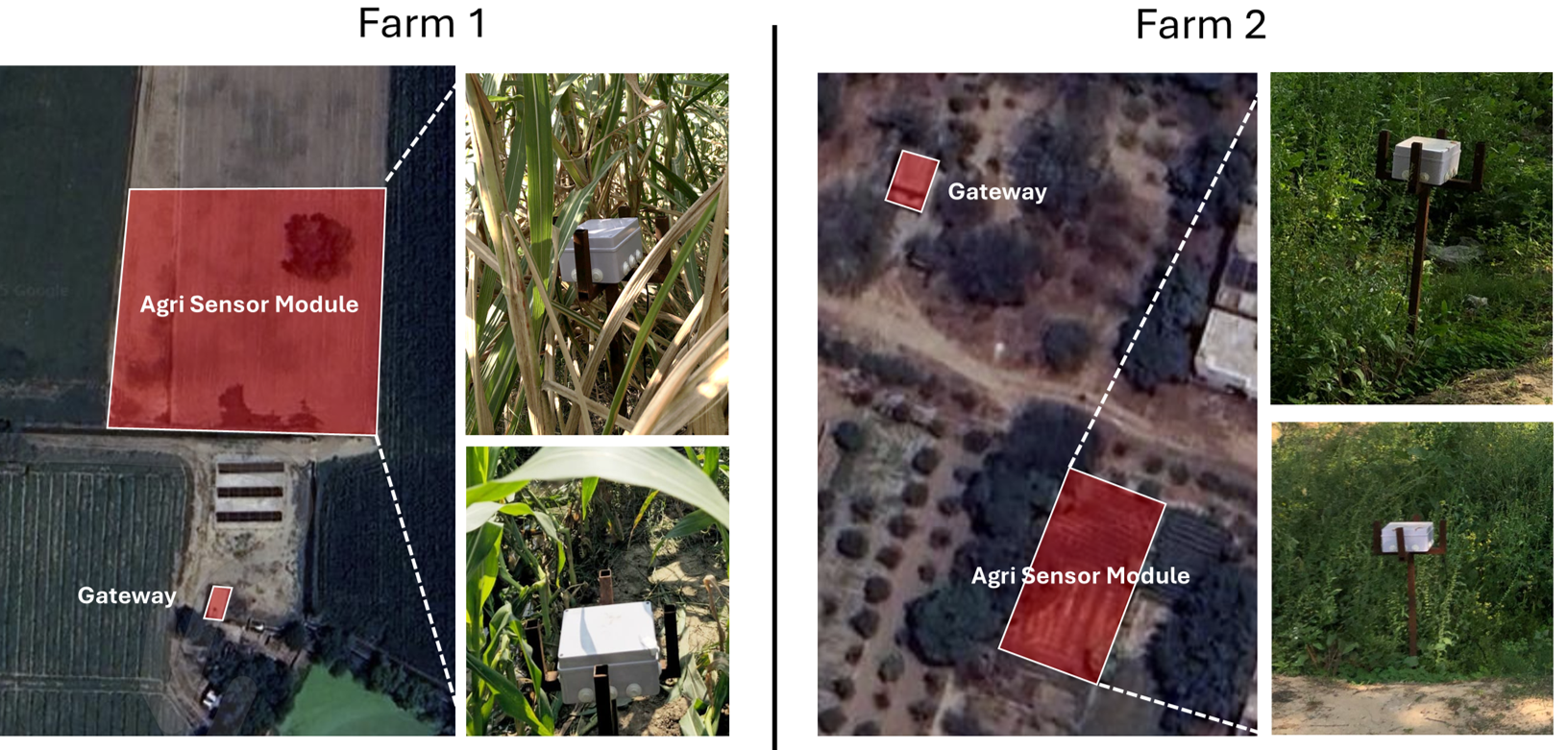}
    }
    \caption{Aerial and field views of the Agri Sensor Module and Gateway installations at Farm 1 (sugarcane) and Farm 2 (vegetable farm).}
    \label{fig:farms}
\end{figure}

To test real-world viability, we ran a 90-day formative pilot study after IRB approval. Rather than a large-scale quantitative trial, we opted for an in-depth, qualitative approach, installing \kd{} at two contrasting Punjab sites: a 2.5-acre commercial sugarcane field and a 0.5-acre university plot that grows organic vegetables for faculty. Five people, two managers and three field workers\footnote{One worker left midway for health reasons.}, used the system for 45 days at each site. Figure~\ref{fig:farms} shows the sensor layouts.

The mix of commercial and academic plots lets us observe both strategic planning and daily field work. This pilot aimed to (i) verify hardware and cloud robustness under field conditions, (ii) gather qualitative feedback on the WhatsApp interface, and (iii) surface challenges and opportunities for a larger-scale study, rather than to produce statistically significant yield outcomes.

\subsection{Methodology}

\fakepara{Deployment phases}  
Each site followed three successive 15-day periods. The opening phase observed existing practices and captured decision-making through baseline interviews\footnote{We released code and documentation via anonymized GitHub repo~\cite{kissan_dost_code}.}. Next, participants used a localized Urdu dashboard (Figure~\ref{fig:dash-graph}) that visualized temperature, conductivity,  NPK, pH, and moisture with trend lines, serving as a visual benchmark. Finally, they switched to the \kd{} WhatsApp chatbot(e.g., Figure~\ref{fig:whatsapp1}), which delivered the same sensor insights in conversational form and in the user’s preferred language.

\fakepara{User onboarding}  
During the baseline interview, we recorded each farmer’s phone number, language, crops, and location, then sent a test message via the WhatsApp Business API. A successful reply activated the account and locked in the chosen language, so all later chats were automatically localised.

\fakepara{Data collection and analysis}  
We combined system logs, field notes, and semi-structured interviews. Interview transcripts were coded thematically with an evolving code-book to track adoption, and behaviour change. Triangulating qualitative insights with usage metrics and observations revealed patterns in accessibility, knowledge gain, and how trust formed across the three phases. Summary of instruments shown in Table \ref{tab:data-collection}.

\begin{table}
\centering
\vspace{0.2cm}
\caption{Data collection methods and analysis}
\label{tab:data-collection}
\resizebox{\columnwidth}{!}{
\begin{tabular}{p{1.5cm}p{2.5cm}p{2.6cm}p{2.9cm}}
\toprule
\textbf{Method} & \textbf{Description} & \textbf{Data Captured} & \textbf{Analysis Approach} \\
\midrule
System Logs & WhatsApp conversations with the chatbot & Query types, usage patterns, interaction frequency & Frequency analysis~\cite{Jung2024} \\
Semi-Structured Interviews & Pre- and post-deployment semi-structured interviews & Perceptions of trust, usability, knowledge gain, adoption barriers & Thematic analysis using structured codebook~\cite{Chakravorti2025} \\
Field Observations & Researcher notes during site visits & Non-verbal behaviors, contextual challenges & Qualitative pattern identification \\
\bottomrule
\end{tabular}}
\end{table}

\subsection{User Experience}
\label{sec:user_study_design}
Our field deployment surfaced insights into decision-making, usability, trust, and knowledge gains across the three study phases.

\fakepara{Phase 1: Baseline agricultural practices}
During baseline, decisions at both farms were largely experience-driven. Irrigation, weeding, and field preparation relied on visual inspection, seasonal heuristics, and peer input, with limited use of external data (e.g., weather forecasts).

Managers were comfortable with smartphones but had little exposure to digital agriculture tools. Generic weather apps were used occasionally but considered unreliable for plot-specific decisions, reinforcing skepticism toward automated advisories.

\begin{figure}[t] 
    \centering 
    \begin{minipage}[t]{0.39\columnwidth} 
        \centering
        \includegraphics[width=\linewidth]{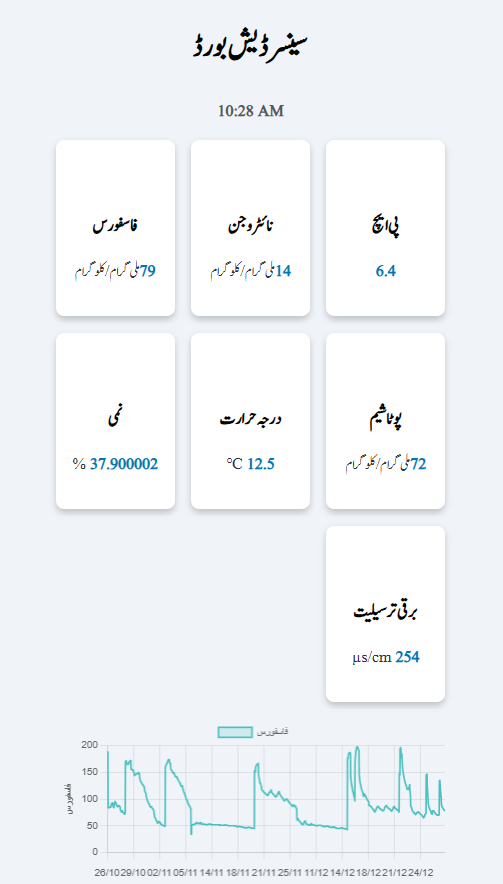} 
        \caption{Dashboard.}
        \label{fig:dash-graph} 
    \end{minipage} 
    \begin{minipage}[t]{0.39\columnwidth} 
        \centering
        \includegraphics[width=\linewidth]{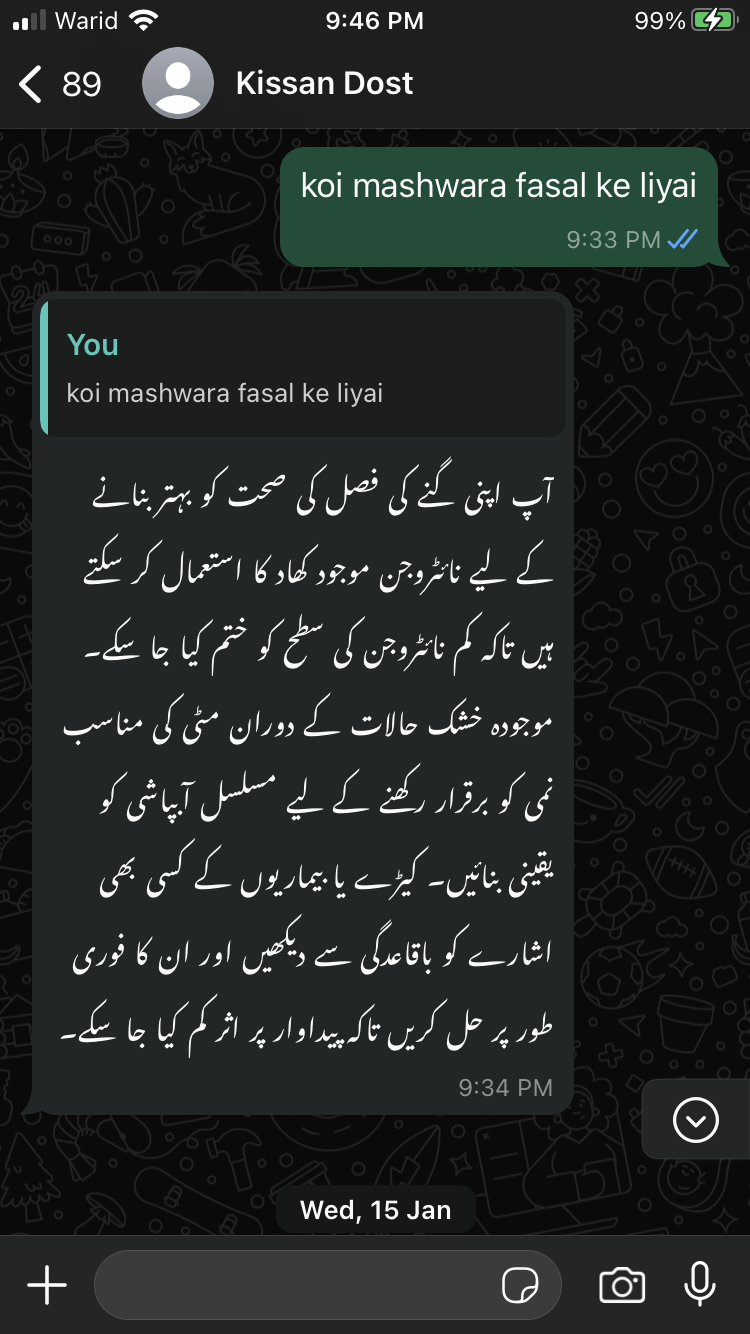} 
        \caption{Chatbot.}
        \label{fig:whatsapp1}
    \end{minipage}
\end{figure}

\begin{figure}[t]
    \centering
    \includegraphics[width=\linewidth]{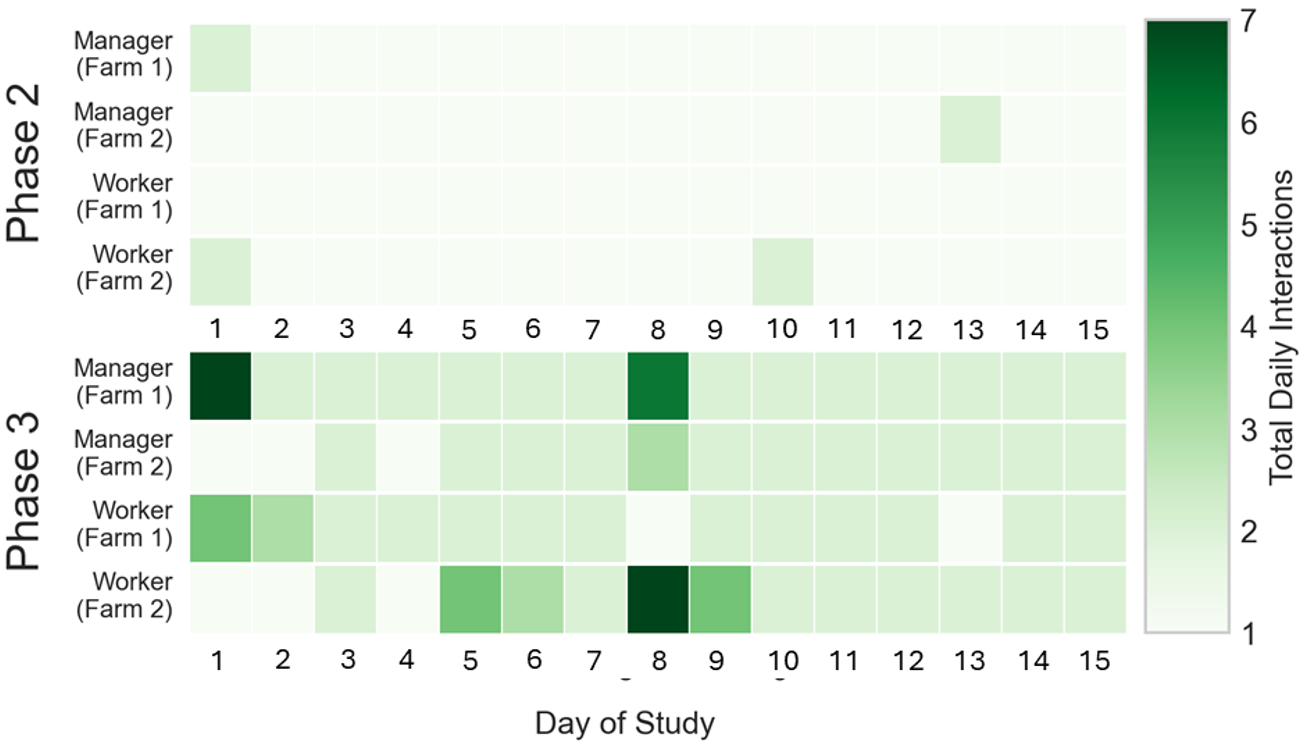}
    \caption{Daily participant interactions during Phase 2 (Dashboard) and Phase 3 (Chatbot) over 15-day periods.}
    \label{fig:chatbot_interaction_trend}
\end{figure}

\fakepara{Phase 2: The dashboard usability challenge}
Introducing the IoT web dashboard (Figure~\ref{fig:dash-graph}) led to consistently low engagement\footnote{Engagement defines the number of times the dashboard app was opened.} across both sites (Figure~\ref{fig:chatbot_interaction_trend}, Phase 2). Managers opened it sporadically but reported that interpreting graphs required agronomic expertise, limiting sustained use. Workers largely disengaged; for example, Worker (Farm 1) never opened the dashboard during the 15-day phase, even after the demonstration. His sentiments\footnote{All comments, originally in local language, were translated into English} were:

\begin{chatlogexcerpt}[blue][Interview Excerpt]  \chatline{
I am unable to understand the dashboard as the random numbers make no sense to me... I would rather make decisions according to my own knowledge rather than deciphering the dashboard.}{Worker, Farm 1}
\end{chatlogexcerpt}

Overall, dashboards increased transparency but did not bridge the gap from raw sensor values to practical decisions for users without technical training.

%%%%%%%%%%%%%%%%%%%%%%%%%%%%%%%%%

\fakepara{Phase 3: Chatbot adoption and engagement}  
Participants quickly abandoned the dashboard but maintained steady WhatsApp use; Figure~\ref{fig:chatbot_interaction_trend} contrasts Phase 3’s activity with Phase 2’s near-silence. This supports our claim: farmers engage when advice is interpretable and delivered in their language. Since agronomic conditions shift over days, success is reflected in sustained daily use rather than message volume; two to three exchanges per day were sufficient. Farm 1 adopted \kd{} immediately, while Farm 2 ramped up after early recommendations proved accurate. By Day 8, the worker at Farm 2 consulted the bot regularly, coinciding with warnings about declining crop health.

\begin{figure}[t] \centering
    \includegraphics[width=\linewidth]{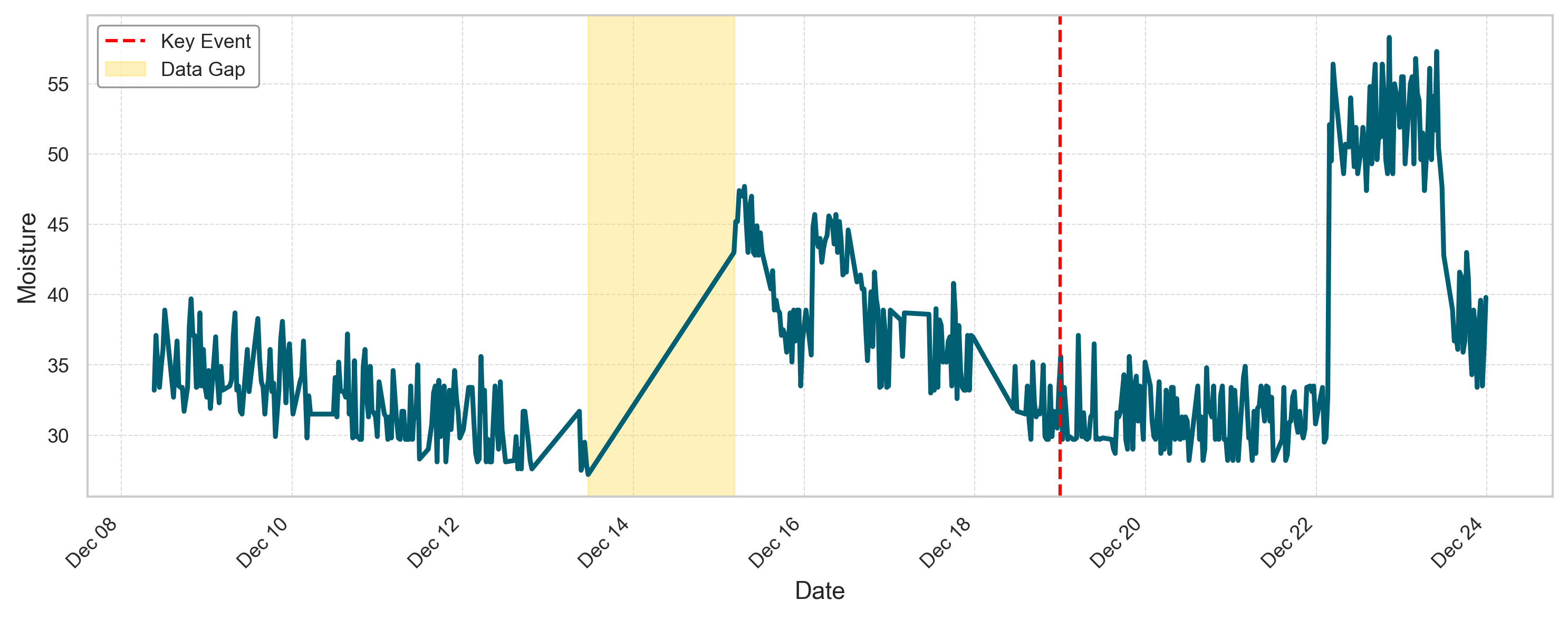}
    \caption{Moisture trends (Commercial Farm, Phase 3). Chatbot recommendation to delay irrigation (red dotted line) aligned with sensor data.}
    \label{fig:moisture_trends}
\end{figure}

\begin{figure}[t] \centering
    \includegraphics[width=\linewidth]{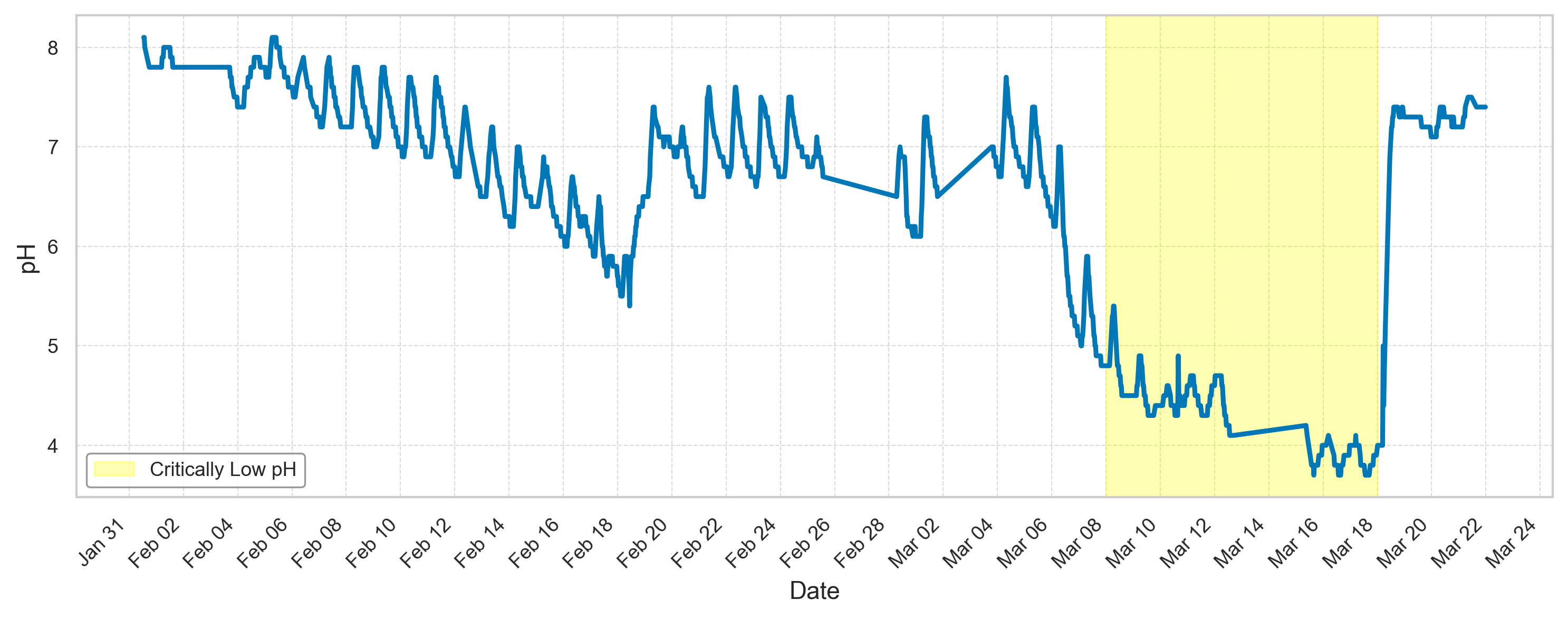}
    \caption{pH trends (University Farm). Critically low pH values recorded by sensors, corresponding to chatbot warnings about acidity and potential disease.}
    \label{fig:Ph}
\end{figure}

Two episodes illustrate impact. At Farm 1, the manager postponed irrigation after the chatbot flagged adequate soil moisture (Figure~\ref{fig:moisture_trends}). At Farm 2, the bot detected a pH drop (Figure~\ref{fig:Ph}), issued repeated alerts, and suggested lime:

\begin{chatlogexcerpt}[green] \chatline{There are some issues with your spinach, particularly regarding pH and nitrogen levels... your soil pH is quite low, around 4.3 to 4.7... consider applying lime or neem oil to increase pH... Nitrogen levels are also below ideal...}{Chatbot}
\end{chatlogexcerpt}

The worker could not afford the treatment but confirmed the diagnosis:

\begin{chatlogexcerpt}[blue][Interview Excerpt] \chatline{All questions I asked were answered correctly. Regarding disease prevention, you mentioned it correctly; although I didn't implement it practically, the suggestions seemed very accurate. When the plant got spotting... it recommended... organic spray (neem oil) or lime but I could not use it, though the advice was solid and correct...}{Worker, Farm 2} 
\end{chatlogexcerpt}
% Figure: Moisture Trends (Keep as is)

Across both farms, users highlighted the simple language, voice-note option, and familiar WhatsApp interface:

\begin{chatlogexcerpt}[blue][Interview Excerpt]  \chatline{The chatbot is easier to use because it explains things and provides written advice.}{Worker, Farm 1}
\end{chatlogexcerpt}

\begin{chatlogexcerpt}[blue][Interview Excerpt]  \chatline{
The information was so clear that even an illiterate person like me can use this kind of technology, which now actually makes sense to me.}{Worker, Farm 2}
\end{chatlogexcerpt}

By translating sensor streams into clear, local-language guidance, \kd{} was both accessible and actionable in day-to-day farming.

% Figure: Ph Trends (Keep as is)

\subsection{Synthesized Insights and Implications}

Analysis across both deployments revealed consistent themes regarding the chatbot’s impact:

\fakepara{Accessibility \& reduced cognitive load}  
The chatbot’s simplified, conversational approach lowered the barrier to interpreting complex sensor data, particularly compared to dashboards. This proved effective for users regardless of literacy level. The familiar WhatsApp interface further eased adoption. Users frequently cited clarity and ease of understanding:

\begin{chatlogexcerpt}[black][Interview Excerpt]  \chatline{An ordinary person who hasn't studied... obviously they're going to need recommendations from the bot.}{Manager, Farm 1}
\end{chatlogexcerpt}

\fakepara{Trust formation}  
Trust was built through the chatbot’s perceived \textit{precision}, such as site-specific, root-level insights, as well as the reinforcement of farmer intuition and relevant contextual cues. Participants appreciated advice aligned with local conditions, including planting times and observable field trends. For instance:

\begin{chatlogexcerpt}[blue][Interview Excerpt]  \chatline{
Whenever I was confused, the chatbot clarified perfectly... it even mentioned local planting timings in Pakistan... it even knew things such as that which made me trust it a little bit.}{Worker, Farm 2}
\end{chatlogexcerpt}

\begin{chatlogexcerpt}[black][Interview Excerpt]  \chatline{So this thing gives information at root depth... it makes a lot of difference.}{Manager, Farm 1}
\end{chatlogexcerpt}

\fakepara{Knowledge expansion \& decision support}  
Participants described gaining new insights into agricultural practices, including organic pest control and optimal fertilization timing. These knowledge gains translated into operational decisions. For example:

\begin{chatlogexcerpt}[blue][Interview Excerpt]  \chatline{
It gave better answers than my knowledge and increased my knowledge greatly.}{Worker, Farm 2}
\end{chatlogexcerpt}

\begin{chatlogexcerpt}[black][Interview Excerpt]  \chatline{You can see what nutrients are lacking....... So we could adjust the application...}{Manager, Farm 1}
\end{chatlogexcerpt}

\fakepara{Limitations \& future needs}  
Participants identified two key limitations. First, the chatbot lacked crop lifecycle awareness, limiting its contextual relevance during different growth stages. Second, the short duration of the study constrained its potential for long-term guidance:

\begin{chatlogexcerpt}[black][Interview Excerpt]  \chatline{The chatbot has to be informed about the stage of the crop}{Manager, Farm 1}
\end{chatlogexcerpt}

\begin{chatlogexcerpt}[blue][Interview Excerpt]  \chatline{
I wanted comprehensive guidance for seasonal vegetables... But I only got 15 days' worth of time.}{Worker, Farm 2}
\end{chatlogexcerpt}

%%%%%%%%%%%%%%%%%%%%%%%%%%%%%%%%%%%%%%%%%%%%%%%%%%%%%%%%%%%%%%%%%%%%%%%%%%%%%5

% \section{Discussion}

% \kd{} highlights that, for smallholders, the limiting factor is rarely sensing itself but the last-mile translation of measurements into decisions. Commodity ESP32 nodes plus basic backhaul were sufficient once advice was delivered through a familiar channel and tied to transparent evidence. Designing an IoT--LLM pipeline surfaced a grounding-versus-synthesis tension: retrieval improves reliability, but practical agronomic guidance often requires inference across sensors, forecasts, and local best practices. Future systems should make this inference explicit (e.g., citing measurements, showing assumptions, or supporting quick follow-up questions) to preserve trust.

% Our pilot also exposed operational constraints, especially intermittent connectivity. Extending \kd{} to areas without reliable backhaul likely requires edge/offline modes. While the study spans only two farms, the interface barriers and trust dynamics we observed are common in LMIC deployments; longer trials across full cropping cycles are needed to measure impacts on productivity and resource use.

\section{Conclusion}

\kd{} shows that combining off-the-shelf soil sensors with retrieval-augmented generation and a WhatsApp chatbot can deliver precision-agriculture advice that smallholder farmers actually use. The system distills live sensor streams and vetted agronomy guidance into clear, multilingual text or voice messages. In controlled tests, it achieved $>$90\% correct sensor-grounded answers, and a 90-day field deployment showed sustained daily engagement and improved irrigation decisions, unlike the dashboard. Overall, the findings highlight that last-mile delivery---language fit, cultural alignment, and trusted channels---is what unlocks the value of existing Agri-IoT hardware in LMIC contexts.

\bibliographystyle{IEEEtran}
\bibliography{sample-base,IEEEabrv}

\end{document}